\documentclass[prl,showpacs,preprint,amssymb]{revtex4}
\usepackage{amsmath,graphicx}
\usepackage{epsfig}

\begin{document}

\title{Polymer desorption under pulling: a novel dichotomic phase
transition}

\author{S. Bhattacharya$^1$, V. G. Rostiashvili$^1$, A. Milchev$^{1,2}$, and
T.A. Vilgis$^1$}
\affiliation{$^1$ Max Planck Institute for Polymer Research, 10 Ackermannweg,
55128 Mainz, Germany\\
$^2$ Institute for Physical Chemistry, Bulgarian Academy of Sciences, 1113
Sofia, Bulgaria
}

\begin{abstract}
We show that the structural properties and phase behavior of a self-avoiding
polymer chain on adhesive substrate, subject to pulling at the chain end, can
be obtained by means of a Grand Canonical Ensemble (GCE) approach. We derive
analytical expressions for the mean length of the basic structural units of
adsorbed polymer, such as  loops and tails, in terms of the adhesive
potential and detachment force, and determine values of the universal exponents
which govern their probability distributions. Most notably, the hitherto
controversial value of the critical adsorption exponent $\phi$ is found to
depend essentially on the interaction between different loops. The chain
detachment transition turns out to be of the first order, albeit dichotomic,
i.e., no coexistence of different phase states exists. These novel theoretical
predictions and the suggested phase diagram of the adsorption-desorption
transformation under external pulling force are verified by means of extensive
Monte Carlo simulations.
\end{abstract}
\pacs{05.50.+q, 68.43.Mn, 64.60.Ak, 82.35.Gh, 62.25.+g}
\maketitle

{\it Introduction} - The manipulation of single polymer chains  has turned
recently into an  important method for understanding their  mechanical
properties and characterization of the intermolecular interactions
\cite{Strick,Celestini}. Such manipulation is mainly triggered  by the progress
in atomic force microscopy (AFM) \cite{Rief} as well as by the development of
optical/magnetic tweezers technique \cite{Bustamante}. This rapid development
has been followed by theoretical considerations, based on the mean - field
approximation \cite{Sevick}, which provide important insight into the mechanism
of polymer detachment from adhesive surfaces under the external force pulling. A
comprehensive study by Skvortsov {\it et al.} \cite{SKB} examines the case of a
Gaussian polymer chain. We also note here the close analogy between the force
detachment of adsorbed chain and the unzipping of a double - stranded DNA.
Recently, DNA denaturation and unzipping have been treated by Kafri {\it et al.}
\cite{Kafri} using the Grand Canonical Ensemble (GCE) approach \cite{Poland} as
well as Duplantier's analysis of polymer networks of arbitrary topology
\cite{Duplantier}. An important result for the properties of adsorbed
macromolecule under pulling turns to be the observation \cite{Kafri} that the
universal exponents (which govern polymer statistics) undergo renormalization
due to excluded volume effects, leading thus to a change of the order of DNA
melting transition from second to first order. In this Letter we use similar
methods to describe the structure and detachment of a single chain from a sticky
substrate when the chain end is pulled by external force.%

{\it Single chain adsorption} - Starting with the conventional (i.e.,
force-free) adsorption,  we recall that an adsorbed chain is build up from
loops, trains, and a free tail. One can treat statistically these basic
structural units by means of the GCE approach where the lengths of the buildings
blocks are not fixed but may rather fluctuate. The GCE-partition function is
then given by
\begin{eqnarray}
 \Xi (z) = \sum_{N=0}^{\infty} \: \Xi_{N} \: z^{N} = \frac{V_{0}(z) \:
Q(z)}{1 - V(z) U (z)},
\label{GC_partition}
\end{eqnarray}
where $z$ is the fugacity and $U(z)$, $V(z)$, and $Q(z)$ denote the GCE
partition functions of loops, trains and tails, respectively. The building block
adjacent to the tethered chain end is allowed for by $V_{0}(z) = 1+V(z)$. The
partition function of the loops is defined as $ U (z)  = \sum_{n=1}^{\infty} \:
(\mu_3 z)^{n}/n^{\alpha}$, where $\mu_3$ is the $3d$ connective constant and
$\alpha$ is the exponent which governs surface loops statistics. It is well
known that for an {\it isolated } loop $\alpha = 1-\gamma_{11} \approx 1.39$
\cite{Vanderzande}. We will argue below that $\alpha$ changes value due to the
excluded volume interaction between a loop and the rest of the chain. The train
GCE-partition function reads $V (z)  = \sum_{n=1}^{\infty} \: (\mu_3 w
z)^{n}/n^{1-\gamma_{d=2}}$ whereby one assumes that each adsorbed segment gains
an additional statistical weight $w = \exp(\epsilon)$ with the dimensionless
adsorption energy $\epsilon = \varepsilon/k_BT$. Eventually, the GCE partition
function for the chain tail is defined by $Q(z) = 1 + \sum_{n=1}^{\infty} \:
(\mu_3 z)^{n} / n^{\beta}$. For an isolated tail $\beta = 1-\gamma_1 \approx
0.32$ \cite{Vanderzande} but again the excluded volume interactions of a tail
with the rest of the chain increase the value of $\beta$. Using the {\em
generating function} method \cite{Rudnick}, $\Xi_N$ is obtained as $\Xi_N =
(z^*)^{-N}$ where the pole $z^*$ is given by the condition $V(z^*) U(z^*) =
1$ so that the free energy is $F=k_BTN \ln z^*$ and the fraction of adsorbed
monomers $n = - \partial \ln z^*/\partial \ln w$. In terms of the so called {\it
polylog function}, which is defined as $\Phi (\alpha, z)=\sum_{n=1}^{\infty} \:
z^{n}/n^{\alpha}$ \cite{Erdelyi}, the equation for $z^*$ reads
\begin{eqnarray}
\Phi(\alpha, \mu_3 z^*) \Phi(\lambda, \mu_2 w z^*) = 1.
\label{Basic_Eq}
\end{eqnarray}
A nontrivial solution for $z^*$ in terms of $w$ (or the adsorption energy
$\epsilon$) appears at the {\em critical adsorption point} (CAP) $w=w_c$ where
$w_c$ is determined from $ \zeta (\alpha)\Phi(1-\gamma_{d=2}, \mu_2
w_c/\mu_3))=1$ as $z^*=1/\mu_3$ and $\zeta(\alpha)$ is the Riemann function. In
the vicinity of the CAP the solution attains the form
\begin{eqnarray}
 z^*(w) \approx  [1 -  A \: (w - w_c)^{1/(\alpha-1)}]\mu_3^{-1}
\label{Solution}
\end{eqnarray}
where $A$ is a constant. Then the average fraction of adsorbed monomers is $n
\propto (\epsilon - \epsilon_c)^{1/(\alpha-1) - 1}$. A comparison with the well
known scaling relationship $n \propto (\epsilon - \epsilon_c)^{1/\phi - 1}$
where $\phi$ is the so called {\it adsorption (or, crossover) exponent}
\cite{Vanderzande} suggests that
\begin{eqnarray}
 \phi = \alpha - 1
\label{Phi}
\end{eqnarray}
This is a result of principal importance. It shows that the crossover exponent
$\phi$, describing polymer adsorption at criticality, is determined by the
exponent $\alpha$ which governs polymer loop statistics! If loops are treated as
isolated objects, then $\alpha=1-\gamma_{11}\approx 1.39$ so that $\phi = 0.39$.
In contrast, excluded volume interactions between a loop and the rest of the
chain lead to an increase of $\alpha$ and $\phi$, as we show below.

{\it Probability  distributions of loops and tails} - How does the length
distribution of polymer loops and tails close to the CAP look like? From the
expression for $U(z)$, given above, and eq.(\ref{Solution}) we have $P_{\rm
loop} \approx (\mu_3 z^*)^l/l^{1+\phi} \approx \exp[ - c_1 (\epsilon -
\epsilon_c)^{1/\phi}] / l^{1+\phi}$. This is valid only for $\epsilon >
\epsilon_c$ since a solution for eq.(\ref{Basic_Eq}) for subcritical values of
the adhesive potential $\epsilon$  does not exist. Nontheless, even in the
subcritical region, $\epsilon < \epsilon_c$, there are still monomers which
occasionally touch the substrate, creating thus single loops at the expense of
the tail length. The partition function of such a loop-tail configuration is
$Z_{l-t} = \frac{\mu_3^l}{l^{1+\phi}}\;\frac{\mu_3^{N-l}}{(N-l)^\beta}$. On the
other side, the partition function of a tail conformation with no loops
whatsoever (i.e., of a nonadsorbed tethered chain) is $Z_t =
\mu_3^N\;N^{\gamma_1-1}$. Thus the probability $P^<_{\rm loop}(l)$ to find a
loop of length $l$ next to a tail of length $N-l$ can be estimated as $
P^<_{\rm loop}(l) = \frac{Z_{l-t}}{Z_t} \propto \frac{N^{1-\gamma_1}}{l^{1+\phi}
(N-l)^{\beta}}$, which is valid at $\epsilon < \epsilon_c$. In the vicinity of
the CAP, $\epsilon \approx  \epsilon_c$, the distribution will be given by an
interpolation between the expressions above. Hence, the overall loop
distribution becomes
\begin{eqnarray}
P_{\rm loop}(l) =  \begin{cases}
                      \frac{1}{l^{1+\phi}}\exp\left [ -c_1 (\epsilon
-\epsilon_c)^{1/\phi}\;l\right ], \quad & \quad \epsilon >
\epsilon_c\\
\frac{A_1}{l^{1+\phi}} + \frac{A_2 N^{1-\gamma_1}}{l^{1+\phi} (N-l)^{\beta}},
\quad & \quad \epsilon = \epsilon_c\\
\frac{N^{1-\gamma_1}}{l^{1+\phi} (N-l)^{\beta}}. \quad & \quad  \epsilon <
\epsilon_c
                  \end{cases}
\label{Loop_distribution__all}
\end{eqnarray}
The same reasonings for a tail leads to the distribution
\begin{eqnarray}
P_{\rm tail}(l) =  \begin{cases}
                      \frac{1}{l^\beta}\exp\left [ -c_1 (\epsilon
-\epsilon_c)^{1/\phi}\;l\right ], \quad & \quad \epsilon >
\epsilon_c\\
\frac{B_1}{l^\beta} + \frac{B_2 N^{1-\gamma_1}}{l^\beta (N-l)^{1+\phi}}, \quad
& \quad \epsilon = \epsilon_c\\
\frac{N^{1-\gamma_1}}{l^\beta (N-l)^{1+\phi}}. \quad & \quad  \epsilon <
\epsilon_c
                  \end{cases}
\label{Tail_distribution__all}
\end{eqnarray}
In eqs.(\ref{Loop_distribution__all}) - (\ref{Tail_distribution__all}) $A_1,
A_2, B_1, B_2$ are constants. Close to CAP these distributions are expected to
attain a U~-~shaped form (with two maxima at $l=1$ and $l \approx N$), as
predicted for a Gaussian chain by Gorbunov {\it et al.} \cite{Gorbunov}. For the
average loop length $L$ the GCE-partition function for loops yields $L=z
\partial U(z)/\partial z |_{z=z^*} = \Phi (\alpha-1, \mu_3 z^*)/\Phi (\alpha,
\mu_3 z^*)$. Close to the CAP, $L$ diverges as $L \propto 1/(\epsilon -
\epsilon_c)^{1/\phi - 1}$. The average tail length $S$ can be obtain as $S=z
\partial Q(z)/\partial z |_{z=z^*} = \Phi (\beta-1, \mu_3 z^*)/[1+ \Phi (\beta,
\mu_3 z^*)]$. Again, using the properties of the polylog function, one can show
that close to $\epsilon_c$  the average tail length diverges as $S \propto
1/(\epsilon - \epsilon_c)^{1/\phi}$. Note that this behavior  corresponds
to a length of adsorption blob $g \propto 1/(\epsilon - \epsilon_c)^{1/\phi}$.

{\em Role of interacting loops and tails} - Consider the number of
configurations of a tethered chain in the vicinity of the CAP as an array of
loops which end up with a tail. Using the approach of Kafri {\it et al.}
\cite{Kafri} along with Duplantier's \cite{Duplantier} graph theory of polymer
networks, one may write the partition function $Z$ for a chain with ${\cal N}$
building blocks: ${\cal N}-1$ loops and a tail. Consider a loop of length $M$
while the length of the rest of the chain is $K$, that is, $M+K=N$. In the limit
of $M \gg 1,\; K\gg 1$ (but with $M/K \ll 1$) one can show \cite{SBVRAMTV}
that $Z \sim \mu_{3}^{M} \: M^{\gamma_{\cal
N}^s - \gamma_{{\cal N}-1}^s} \:\: \mu_{3}^{K} \: K^{\gamma_{{\cal N}-1}^s - 1}$
where the surface exponent $ \gamma_{\cal N}^s = 2  - {\cal N}(\nu +
1) + \sigma_1 + \sigma_1^s$ and $\sigma_1,\; \sigma_1^{s}$ are critical bulk and
surface exponents \cite{Duplantier}. The last result indicates that the
effective loop exponent $\alpha$ becomes
\begin{equation}
 \alpha = \gamma_{{\cal N}-1}^s - \gamma_{\cal N}^s = \nu + 1
\label{Effective_alpha}
\end{equation}
Thus, $\phi=\alpha-1=\nu = 0.588$, in agreement with earlier Monte Carlo
findings \cite{Eisenriegler}. One should emphasize, however, that the foregoing
derivation is Mean-Field-like ($Z$ appears as a product of loop- and
rest-of-the-chain contributions) which overestimates
the interactions and increases significantly the value of $\alpha$, serving as
an upper bound. The value of $\alpha$, therefore, is found to satisfy
the inequality $1-\gamma_{11}\le \alpha \le 1+\nu$, i.e., depending on loop
interactions, $0.39 \le \phi \le 0.59$.

{\it Adsorption under detaching force} -
Using the GCE approach now we treat the case of self-avoiding polymer
chain adsorption in the presence of pulling force, thus extending the
consideration of Gaussian chains by Gorbunov {\it et al.} \cite{Skvortsov}.
Under external detaching force $f$, the tail GCE-partition function $Q(z)$ in
eq.~(\ref{GC_partition}) has to be replaced by
$\tilde{Q}(z) = 1 + \sum_{n=1}^{\infty} [(\mu_3 z)^n / n^{\beta}] \: \int d^3 r
P_n({\bf r}) \exp(f r_{\perp}/T)$ where $P_n({\bf r})$ is the end-to-end
distance distribution function for a self-avoiding chain \cite{DesCloizeaux}.
After some straightforward calculations the tail GCE-partition function can be
written as
\begin{eqnarray}
\tilde{Q}(z) = 1 + a_1 \:{\tilde f}^{\theta} \: \Phi (\psi, z \mu_3 \exp (a_2
{\tilde f}^{1/\nu}))
\label{GC_Tail_Force}
\end{eqnarray}
Here the dimensionless force ${\tilde f}=fa/k_BT$, the exponents $\psi=1-\nu$,
and $\theta=(2+t-3\delta/2)/(\delta-1)$ with $t=(\beta-3/2+3\nu)/(1-\nu)$ and
$\delta=1/(1-\nu)$. The function $\tilde{Q}(z)$ has a branch point at
$z^{\#}=\mu_3^{-1} \exp(-a_2 {\tilde f}^{1/\nu})$, i.e., $\tilde{Q}(z) \sim
1/(z^{\#}-z)^{1-\psi}$. One may, therefore, conclude that the total
GCE-partition function $\Xi(z)$ has two singularities on the real axis: the pole
$z^*$, and the branch point $z^{\#}$. It is known (see, e.g., Sec. 2.4.3. in
\cite{Rudnick}) that for $N >> 1$ the main contributions to $\Xi_N$ come from
the pole and the branch singular points, i.e.,
\begin{eqnarray}
 \Xi_N \sim C_1 \: (z^{*})^{-N} + \frac{C_2}{\Gamma(1-\psi)} \: N^{-\psi} \:
(z^{\#})^{-N}
\label{Pole_Branch}
\end{eqnarray}
Thus, for large $N$ only the smallest of these points matters.  On the other
hand, $z^{*}$ depends on the dimensionless adsorption energy $\epsilon$ only
(or, on $w=\exp(\epsilon)$) whereas $z^{\#}$ is controlled by the external force
${\tilde f}$. Therefore, in terms of the two {\it control parameters},
$\epsilon$ and ${\tilde f}$, the equation $ z^{*}(\epsilon) = z^{\#}({\tilde
f})$ determines the critical line of transition between the adsorbed phase and
the force-induced desorbed phase . In the following this line will be called
{\it detachment line}. Below it, $f < f_D$, or above, $f > f_D$, either $z^*$ or
$z^\#$, respectively, contribute to $\Xi_N$. The controll parameters,
$\epsilon_D$ and ${\tilde f}_D$, which satisfy  this equation,  denote
detachment energy and detachment force, respectively. On the detachment line the
system undergoes a {\it first-order phase transition}. The detachment line
itself terminates for ${\tilde f}_D \rightarrow 0$ in the CAP, $\epsilon_c$,
\begin{figure}[htb]
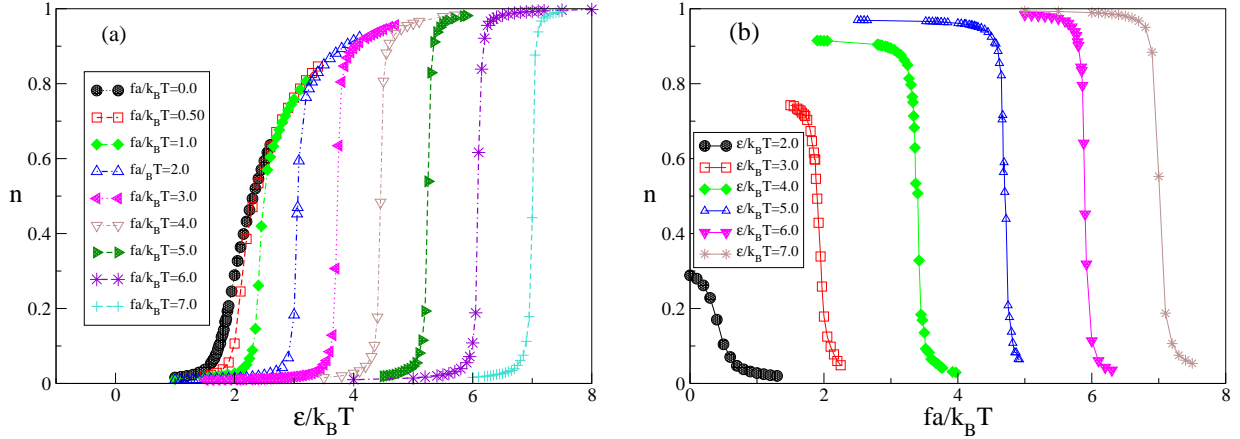

\includegraphics[scale=0.33]{n_e.eps}
\hspace{.3cm}
\includegraphics[scale=0.33]{n_f.eps}
\caption{(a) The 'order parameter', $n$, against the surface potential,
$\epsilon$, for various pulling forces. The chain has length $N$=128. (b)
Variation of $n$ with the pulling force, $f$, for several surface potentials.
}
\label{Order-Parameter}
\end{figure}
where the transition becomes of {\em second} order. In the vicinity of the CAP
the detachment force ${\tilde f}_D$ vanishes as ${\tilde f}_D \sim (\epsilon -
\epsilon_c)^{\nu/\phi}$. This first order adsorption-desorption phase transition
under pulling force has a clear {\em dichotomic} nature (i.e., it follows an
``either - or'' scenario): in the thermodynamic limit $N \rightarrow \infty$
there is {\em no phase coexistence}! The configurations are divided into
adsorbed and detached (or stretched) dichotomic classes. Metastable states are
completely absent. Moreover, the mean loop length $L$ remains finite upon
detachment line crossing. The average tail length $S$, on the contrary, diverges
close to the detachment line. Indeed, at ${\tilde f} < {\tilde f}_D$ the average
tail length is given by  $S={\tilde f}^\theta \Phi (\psi - 1,
z^*(w)/z^{\#}({\tilde f}))/[1+a_1 \Phi (\psi, z^*(w)/z^{\#}({\tilde f}))]$. At
the detachment line, $z^*=z^{\#}$, it diverges as $S \propto {\tilde
f}_D/({\tilde f}_D - {\tilde f})$.
\begin{figure}[htb]
\includegraphics[scale=0.6]{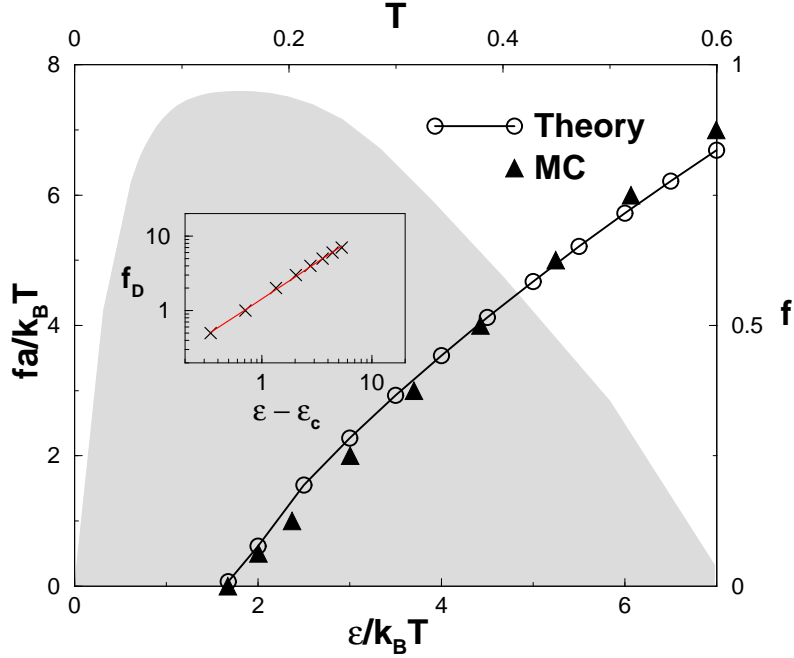}
\caption{Plot of the critical detachment force $f_D=fa/k_BT$ against the surface
potential $\varepsilon/k_BT$. Full and empty symbols denote MC and theoretical
results. A double logarithmic plot of $f_D$ against $\epsilon - \epsilon_c$
with $\epsilon_c = 1.67$ is shown in the inset, yielding a slope of $0.97\pm
0.02$, in agreement with the prediction $f_D \propto
(\epsilon-\epsilon_c)^{\nu/\phi}$. Shaded is shown the same phase diagram,
derived by numeric solution of $z^* = z^\#$, which in dimensional $f$
(right axis) against $T$ (top axis) units appears {\em reentrant}.}
\label{Phase-diagram}
\end{figure}

{\it Reentrant phase behavior} - Recently, it has been realized \cite{Mishra}
that the detachment line, when represented in terms of {\em dimensional}
variables, force $f_D$ versus temperature $T$, goes (at a relatively low
temperature) through a maximum, that is, the desorption transition shows
reentrant behavior! Below we demonstrate that this result follows directly from
our theory. It can be seen that the solution of eq.(\ref{Basic_Eq}) at large
values of $\epsilon$ (or, at low temperature) can be written as $z^* \approx
{\rm e}^{- \epsilon}/\mu_3$ so that the detachment line, $z^{*}=z^{\#}$, in
terms of {\it dimensionless} parameters is monotonous, ${\tilde f}_D  \propto
[\epsilon_D-\ln(\mu_3/\mu_2)]^{\nu}$. Note, however, that the same detachment
line, if represented in terms of the {\it dimensional} control parameters, force
$f_D$ versus temperature $T_D$ (with a fixed dimensional energy
$\varepsilon_0$), shows a nonmonotonic behavior $f_D = T_D [\varepsilon_0/T_D  -
 \ln(\mu_3/\mu_2)]^{\nu}/a$. This curve has a maximum at a temperature
given by $T_{D}^{max} = (1-\nu)\varepsilon_0/\ln(\mu_3/\mu_2) $.

{\it Monte Carlo Simulation} - We have investigated the force induced desorption
of a polymer by means of extensive Monte Carlo simulations using a coarse
grained off-lattice bead-spring model \cite{MC_Milchev} of a polymer.
Fig.~\ref{Order-Parameter}a shows the variation of the order parameter $n$
(average fraction of adsorbed monomers) with changing adhesive potential
$\epsilon$ at fixed pulling force whereas Fig.\ref{Order-Parameter}b depicts $n$
vs. force $f a/T$ for various $\epsilon$. The abrupt change of the order
parameter is in close agreement with our theoretical prediction. Using the
threshold values of $f_D$ and $\epsilon_D$ for critical adsorption/desorption in
the thermodynamic limit $N\rightarrow \infty$, one can construct the adsorption
- desorption phase diagram for a polymer chain under pulling shown in
Fig.\ref{Phase-diagram} which is among the central results of this work. The
detachment lines, obtained from MC data and the numerical solution of $z^* =
z^\#$ almost coincide, and the slope of $f_D$ vs $(\epsilon - \epsilon_c)$ is
close to unity, according to the prediction $f_D \propto
(\epsilon-\epsilon_c)^{\nu/\phi}$. Also indicated by the shaded area in
Fig.\ref{Phase-diagram} is the {\em reentrant} image of the same phase diagram,
obtained when the numerical solution of $z^* = z^\#$ is plotted in dimensional
units $f$ versus $T$.
\vspace{0.5cm}
\begin{figure}[bht]
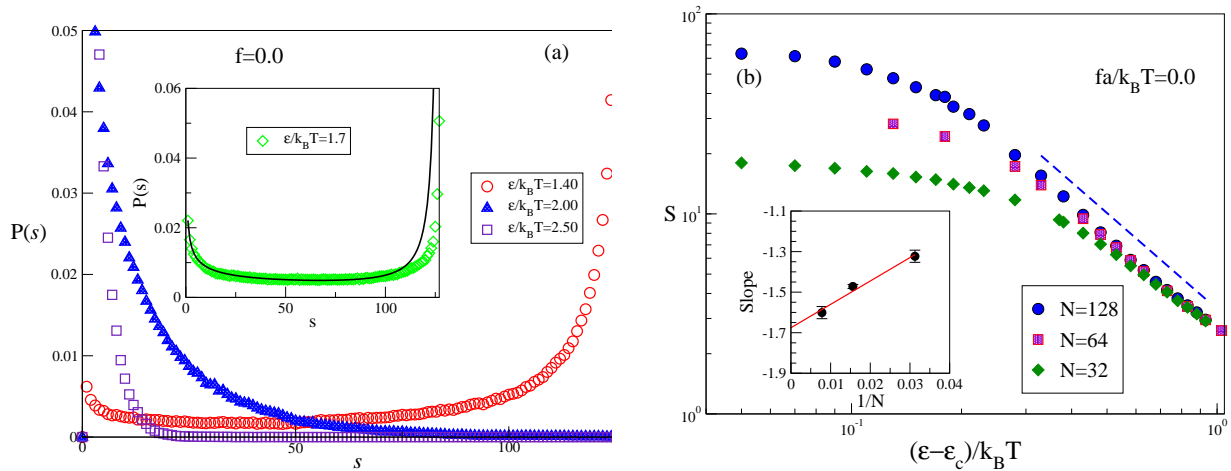

\includegraphics[scale=0.33]{PDF_s.eps}
\hspace{.1cm}
\includegraphics[scale=0.33]{S_e.eps}
\caption{(a) Tail length distribution $P(s)$ for different surface potentials
close to $\epsilon_c$ in  a polymer of length $N = 128$ with no pulling force.
In the inset $P(s)$ at $\epsilon = \epsilon_c$ (symbols) is compared to the
prediction, Eq.~(\ref{Tail_distribution__all}) (full line). (b) Average tail
length $S$ against $(\epsilon -\epsilon_c)/k_BT$ plotted for various chain
lengths in log-log coordinates. The slopes of these curves are plotted against
$1/N$ in the inset and extrapolate to $1/\phi$ in the thermodynamic limit
$N\rightarrow \infty$.}
\label{tail_f0}
\vspace{0.5cm}
\end{figure}
In Fig.~\ref{tail_f0}a we show the PDF of tail length at different strength of
adsorption in the absence of pulling. This confirms the U - shape of $P(s)$
predicted by eq.(\ref{Tail_distribution__all}). While for
$s\rightarrow 1$ the agreement with eq.~\ref{Tail_distribution__all} is perfect,
for $s\rightarrow N$ long tails are slightly overestimated by
eq.(\ref{Tail_distribution__all}). This small discrepancy reflects the
dominance of our ``single loop \& tail'' approximation - multiple loops would
effectively reduce the tail size. Fig.~\ref{tail_f0}b shows the divergency of
$S$ close to the critical point
$\epsilon_c$. For chain of finite length $N$, the tail length divergence at
$\epsilon \rightarrow \epsilon_c$ is replaced by a rounding into a plateau since
$S \rightarrow N$ but away from $\epsilon_c$ the measured slope extrapolates to
the theoretical prediction $S\propto 1/(\epsilon-\epsilon_c)^{1/\phi}$. In the
presence of pulling force one observes a remarkable feature of the order
parameter probability distribution
\begin{figure}[bht]
\vspace{0.8cm}
\includegraphics[scale=0.46]{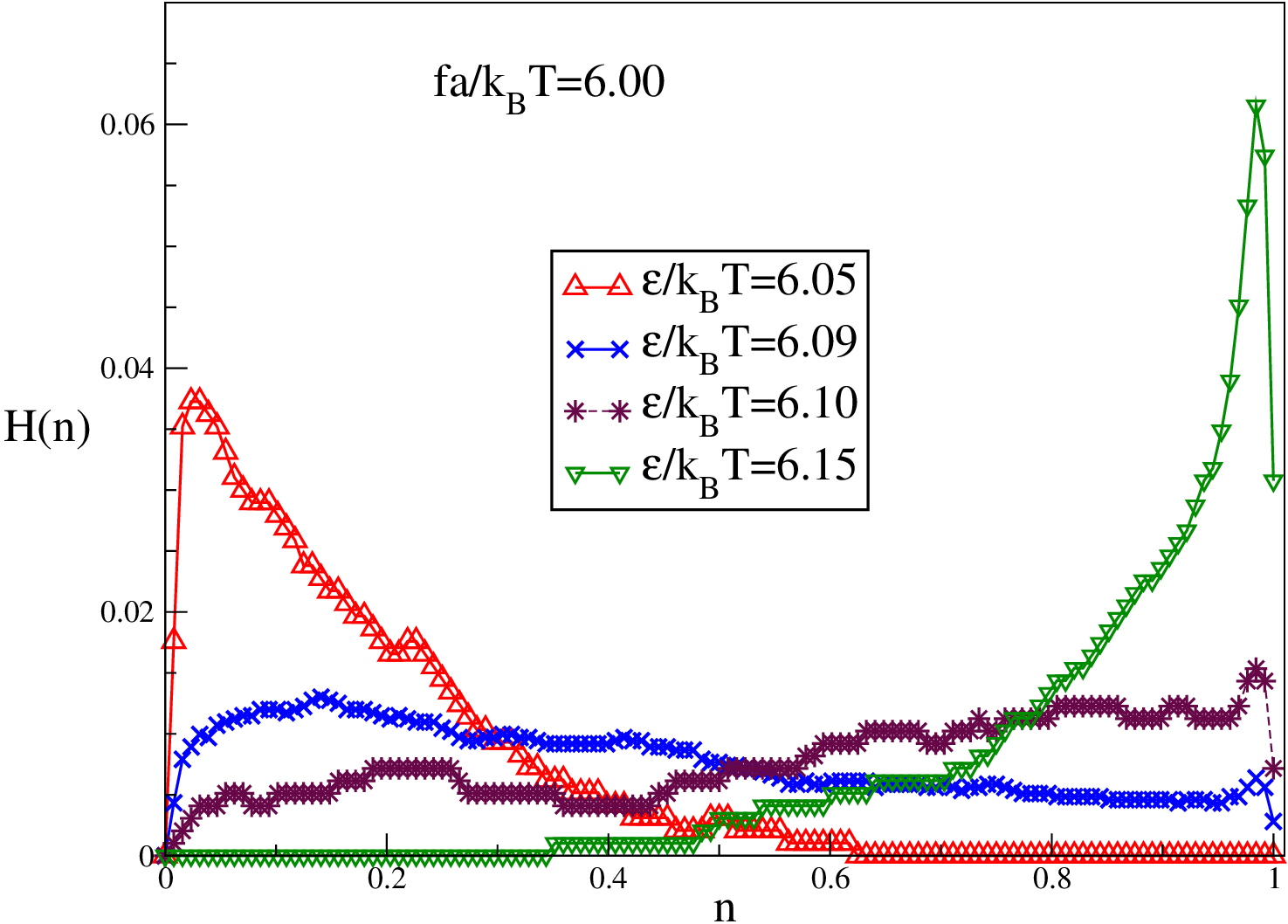}
\caption{Distribution of the order parameter $n$ for a pulling force $fa/k_BT =
6.0$ an different strengths of adhesion $\epsilon/k_BT$. The chain length is
$N=128$ and the threshold value of the surface potential for this force is
$\epsilon_D \approx 6.095 \pm 0.03$. The values $\epsilon/k_BT=6.09$ and
$\epsilon/k_BT=6.10$ are on both sides of the detachment line, cf.
Fig.~\ref{Phase-diagram}.}
\label{fig_PDF_n}
\end{figure}
- an absence of two peaks in the vicinity of the critical strength of
adsorption, $\epsilon_D \approx 6.095 \pm 0.03$, which still keeps the polymer
adsorbed at pulling force $fa/k_BT = 6.0$ - Fig.~\ref{fig_PDF_n}. At
$\epsilon_D$ the distribution $H(n)$ is flat, indicating huge fluctuations so
that any value of $n$ is equally probable.  Close to $\epsilon_D$, one observes
a clear maximum in the distribution $H(n)$, indicating a desorbed chain with
$n\approx 0.01$ for $\epsilon = 6.05$, or a completely adsorbed chain with
$n\approx 0.99$ for $\epsilon = 6.15$. This lack of bimodality in the $H(n)$
manifests the dichotomic nature of the desorption transition which rules out
phase coexistence.

In conclusion, we have demonstrated that a full description of the force
induced polymer chain desorption transition can be derived by means of the GCE
approach, yielding the average size and probability distribution functions of
all basic structural units as well as their variation with changing force or
strength of adhesion. The detachment transition is proved to be of first order
albeit dichotomic in nature thus ruling out phase coexistence. The critical line
of desorption, while monotonous when plotted in dimensionless units of
detachment force against surface potential, becomes non-monotonous in units of
force against temperature, thus outlining a reentrant phase diagram. In addition,
we show that the crossover exponent, $\phi$, governing polymer behavior at
criticality, depends essentially on interactions between different loops so that
$0.39 \le \phi \le 0.59$. All these predictions appear in very good agreement
with our Monte Carlo computer simulation results.

{\em Acknowledgments}
We are indebted to A. Skvortsov, L. Klushin, J.-U. Sommer, and K. Binder for
useful discussions during the preparation of this work. A.~Milchev thanks the
Max-Planck Institute for Polymer Research in Mainz, Germany, for hospitality
during his visit in the institute. A.~Milchev and V.~Rostiashvili acknowledge
support from the Deutsche Forschungsgemeinschaft (DFG), grant No. SFB 625/B4.

\end{document}